\documentclass{pasj00}
\begin{document}
\SetRunningHead{Author(s) in page-head}{Running Head}
\Received{2008/07/31}%{yyyy/mm/dd}
\Accepted{2008/08/28}%{yyyy/mm/dd}

\title{{\it Suzaku} Observations of SGR\,1900$+$14 and SGR\,1806$-$20}

%%% begin:list of authors
\author{
  Yujin E. \textsc{Nakagawa}\altaffilmark{1,2}
  Tatehiro \textsc{Mihara}\altaffilmark{1}
  Atsumasa \textsc{Yoshida}\altaffilmark{2}
  Kazutaka \textsc{Yamaoka}\altaffilmark{2}\\
  Satoshi \textsc{Sugita}\altaffilmark{2}
  Toshio \textsc{Murakami}\altaffilmark{3}
  Daisuke \textsc{Yonetoku}\altaffilmark{3}
  Motoko \textsc{Suzuki}\altaffilmark{4}\\
  Motoki \textsc{Nakajima}\altaffilmark{5}
  Makoto \textsc{Tashiro}\altaffilmark{6}
  and
  Kazuhiro \textsc{Nakazawa}\altaffilmark{7}
}
\altaffiltext{1}{Institute of Physical and Chemical Research (RIKEN), 2-1 Hirosawa, Wako, Saitama 351-0198}
\email{yujin@crab.riken.jp}
\altaffiltext{2}{Graduate School of Science and Engineering, Aoyama Gakuin University, 5-10-1 Fuchinobe, \\Sagamihara, Kanagawa 229-8558}
\altaffiltext{3}{Department of Physics, Kanazawa University, Kakuma, Kanazawa, Ishikawa 920-1192}
\altaffiltext{4}{Tsukuba Space Center, 2-1-1 Sengen, Tsukuba, Ibaraki 305-8505}
\altaffiltext{5}{College of Science and Technology, Nihon University, 1-8-14 Kanda-Surugadai, \\Chiyoda-ku, Tokyo 101-0062}
\altaffiltext{6}{Department of Physics, Saitama University, 255 Shimo-Okubo, Sakura-ku, Saitama, \\Saitama 338-8570}
\altaffiltext{7}{Department of Physics, University of Tokyo, 7-3-1 Hongo, Bunkyo-ku, Tokyo 113-0033}
%%% end:list of authors

%% `\KeyWords{}' always has to be placed before `\maketitle'.
\KeyWords{stars: pulsars individual(SGR\,1900$+$14, SGR\,1806$-$20)} %Do NOT move this preamble from here!

\maketitle

\begin{abstract}
Spectral and timing studies of {\it Suzaku} ToO observations
of two SGRs, 1900$+$14 and 1806$-$20, are presented.
The X-ray quiescent emission spectra were well fitted
by a two blackbody function or a blackbody plus a power law model.
The non-thermal hard component discovered by INTEGRAL was
detected by the PIN diodes and its spectrum was
reproduced by the power law model reported by INTEGRAL.
The XIS detected periodicity
$P = 5.1998\pm0.0002$\,s for SGR\,1900$+$14 and $P = 7.6022\pm0.0007$\,s for SGR\,1806$-$20.
The pulsed fraction was related to the burst activity for SGR\,1900$+$14.
\end{abstract}

%%%%%%%%%%%%%%%%%%%%%%%%%%%%%%%%%%%%%%%
\section{Introduction}
Magnetars (neutron stars
with a super strong surface magnetic field
$B \sim 10^{15}$\,G -- e.g., \cite{duncan1992, paczynski1992}) have received considerable
attention recently.
The magnetar's field strength exceeds the critical field
$B_{\rm c} \equiv m_{e}^{2}c^{3}/c\hbar \approx 4.4 \times 10^{13}$\,G,
and hence the non-linear effects of quantum electrodynamics
must be considered in the processes of radiation transfer.
Several studies have now found X-ray
sources which are magnetar candidates,
namely the soft gamma repeaters (SGRs) and anomalous X-ray pulsars (AXPs).
They display X-ray quiescent emission with luminosities
$L = 10^{33}$-$10^{35}$\,erg\,s$^{-1}$ (e.g., \cite{woods2006}).
The SGRs and a few AXPs
also exhibit ``short bursts'' with typical durations
$\Delta{t} \sim 100$\,ms and super-Eddington luminosities
$L \sim 10^{40}$\,erg\,s$^{-1}$.
One of the most remarkable burst phenomena
is giant flares from SGRs.
Three SGRs have emitted giant flares over the past three decades.
The most energetic
one had $L \sim 5 \times 10^{47}$\,erg\,s$^{-1}$,
from SGR\,1806$-$20 on 27 December 2004 \citep{terasawa2005}.
The giant flares have been proposed as candidates for some
short gamma-ray bursts (GRB) 
with durations $\Delta{t} \lesssim 2$\,s
(e.g., \cite{hurley2005}).
Delayed X-ray and radio emission following giant flares
or short bursts  display
similar behavior to GRB afterglows
\citep{frail1999, feroci2003, cameron2005, nakagawa2008},
which may also support this hypothesis.

Two spectral models,
a two blackbody function (2BB)
and a blackbody plus a power law model (BB$+$PL),
have been suggested for the quiescent X-ray
emission of the magnetar candidates.
A comprehensive study of many short SGR bursts detected by
the High Energy Transient Explorer 2 (HETE-2)
has shown that 2BB
is preferable, although it is an empirical model \citep{nakagawa2007a}.
Although a physical model for the energy spectra remains controversial
despite observations by several satellites,
strong linear correlations between the lower and higher blackbody temperatures
($kT_{\mathrm{LT}}$, $kT_{\mathrm{HT}}$), and between
the lower and higher blackbody radii ($R_{\mathrm{LT}}^2$, $R_{\mathrm{HT}}^2$)
were found, irrespective of the activity states (i.e., the X-ray quiescent emission
or the short bursts),
using 2BB \citep{nakagawa2007b}.
There are various theoretical models which reproduce
the spectra of the quiescent X-ray emission (e.g., \cite{perna2001, guver2006}).
Even more exotic models have been proposed,
such as that invoking a p-star \citep{cea2006}.

One piece of evidence in favor of the magnetar model
is that magnetic pressure confines the photon-pair plasma from
the giant flare in a (rather) small volume
close to the neutron star \citep{thompson1995}.
One complication is the fact that there has not been
a secure direct measurement of the magnetic field of a magnetar candidate.
One possible such measurement is
an absorption feature interpreted to
be due to proton cyclotron resonance scattering
which has been seen in a spectrum of the precursor emission
of a $\sim1.5$\,s burst from SGR\,1806$-$20 \citep{ibrahim2002}.
This kind of feature can also be seen in the spectrum of
X-ray quiescent emission from AXP\,1RXS\,J170849$-$400910 \citep{rea2003}.
The lack of this feature in most magnetar candidates may be explained by
a model which smears
the absorption feature due to multiple cyclotron resonant
scatterings in a stellar atmosphere and its magnetosphere \citep{guver2006}.

Several studies have reported periods $P = 5$-$13$\,s and
period derivatives $\dot{P} = 10^{-11}$-$10^{-13}$\,s\,s$^{-1}$
for the magnetar candidates \citep{woods2006}.
It is known that
pulse characteristics such as the pulsed fraction and pulse shape
vary in time (e.g., \cite{woods2007}), and involve complex pulse morphology.
However, little has been done to elucidate the underlying physics
of the pulse characteristics.
A recent theoretical study suggests that the pulse characteristics
may be explained by trapped fireballs produced by starquakes \citep{jia2008}.

This paper reports the broadband spectroscopy
of two SGRs, 1900$+$14 and 1806$-$20, observed by {\it Suzaku}.
The relation between pulsed fraction and burst rate will also be discussed.
The distances are assumed to be 14.5\,kpc for SGR\,1900$+$14
\citep{hurley1999, vrba2000} and 8.7\,kpc for SGR\,1806$-$20
\citep{corbel1997, corbel2004, cameron2005, mcclure2005, bibby2008}.

%%%%%%%%%%%%%%%%%%%%%%%%%%%%%%%%%%%%%%%
\section{Observations}
{\it Suzaku} \citep{mitsuda2007} has the capability of giving high quality
broadband spectra of astrophysical X-ray and gamma-ray sources,
because of the great sensitivity of the X-ray imaging spectrometer
(XIS; 0.2-12\,keV; \cite{koyama2007})
and an extremely low background level of the hard X-ray detector
(HXD; 10-700\,keV; \cite{takahashi2007}).
The XIS consists of one back-illuminated (BI) CCD and three
front-illuminated (FI) CCDs.
The HXD consists of PIN diodes (10-70\,keV) and GSO units (40-600\,keV).
The XIS has an effective area of 370\,cm$^{2}$ (BI) and 330\,cm$^{2}$ (FI) at 8\,keV.
Although the effective areas are comparable to the EPIC on-board XMM-{\it Newton},
XIS has better energy resolution below 1\,keV: $\sim$50\,eV (BI) and $\sim$40\,eV (FI) at 0.525\,keV.
The HXD PIN also has better energy resolution, $\sim$3\,keV,
compared with FREGATE\footnote{See http://space.mit.edu/HETE/fregate.html.} on-board HETE-2
(5\,keV at 20\,keV) and
IBIS on-board INTEGRAL (7\,keV at 100\,keV; \cite{winkler1999}).
{\it Suzaku} can observe objects in the energy range 0.2-600\,keV
which covers a lower X-ray band
compared with HETE-2 (2-400\,keV; \cite{ricker2003})
and INTEGRAL IBIS (15\,keV-10\,MeV; \cite{winkler1999}).
Since the HXD field of view is narrow ($\timeform{34'} \times \timeform{34'}$),
it has good sensitivity to weak sources.
The sensitivity of the HXD is $3 \times 10^{-6}$\,photons\,s$^{-1}$\,keV$^{-1}$\,cm$^{-2}$
at 20\,keV \citep{takahashi2007}, which is better than
HETE-2 FREGATE ($\sim 10^{-3}$\,photons\,s$^{-1}$\,keV$^{-1}$\,cm$^{-2}$ at 100\,keV;
\cite{atteia2003})
and INTEGRAL IBIS\footnote{See http://integral.esa.int/integ\_payload\_imager.html.}
($10^{-5}$\,photons\,s$^{-1}$\,keV$^{-1}$\,cm$^{-2}$ at 20\,keV).

When the SGRs 1900$+$14 and 1806$-$20 displayed
intense bursting activity, scheduled target of opportunity (ToO) observations by {\it Suzaku}
were performed.
Table \ref{tab:obs_summary} shows a summary of the observations.

SGR\,1900$+$14 began a very high burst activity on 29 March 2006,
and {\it Swift} recorded more than 40 bursts on that day \citep{israel2008}.
The {\it Suzaku} ToO observation started at
8:43 and ended at 21:52 on 1 April 2006 (UT).
The X-ray quiescent emission was detected by the XIS,
but there was no significant emission in the HXD data.
No burst was found in the XIS and HXD light curves.
There were no {\it Swift} bursts during the {\it Suzaku} observation
(the bottom right panel in figure \ref{fig:1900_tvar}).

A bright long burst from SGR\,1806$-$20
was detected by Konus-Wind and {\it Suzaku}-WAM
on 26 March 2007 \citep{golenetskii2007}.
The {\it Suzaku} ToO observation of this source
started at 15:08 on 30 March 2007 and ended at 1:30 31 March 2007 (UT).
Both the X-ray quiescent emission and the non-thermal hard X-ray emission discovered
by INTEGRAL \citep{molkov2005} were detected by the XIS and the HXD.
Two dim short bursts were detected by {\it Suzaku}.
One of them was a short burst localized by {\it Swift} at 16:14:38,
and the other was detected only by {\it Suzaku} at 17:34:00.
Since these bursts are too weak to perform spectral analyses,
we do not consider them further here.

%%%%%%%%%%%%%%%%%%%%%%%%%%%%%%%%%%%%%%%
\section{Analysis}
\subsection{Data Reduction}
The reduction of both the XIS and HXD event data (v2.0) were
made using HEAsoft\,6.4.1 software.
The latest calibration database (CALDB:\,20080616)
was applied to the unfiltered XIS event data using {\it xispi} (v2008-04-10).
Then the new data were filtered using the basic
criteria\footnote{The XIS basic criteria were derived from http://suzaku.gsfc.nasa.gov/docs/suzaku/analysis/abc.}
and with a grade selection ``GRADE = (0,2-4,6)''
using {\it xselect} (v2.4a).
After that, hot and flickering pixels were removed using {\it cleansis} (v1.7).
Telemetry saturated time intervals were estimated by {\it xisgtigen} (v2007-05-14),
and the time intervals were removed using {\it xselect}.
The sizes of the source and background regions are summarized in table \ref{tab:obs_summary}.
The background regions of the FI CCDs for the SGR\,1806$-$20 observation
are selected near the edge opposite to the source region
(because the object fell on the edge),
while those of the other CCDs were selected near both edges.
Light curves and spectra were extracted from the clean XIS event data
using {\it xselect}.
The response matrix files were created by {\it xisrmfgen} (v2007-05-14) and
the ancillary response function files were generated by {\it xissimarfgen} (v2008-04-05).
The spectra were binned to at least 50 counts in each spectral bin by {\it grppha}.

The barycentric correction was applied to the cleaned XIS event data
by use of {\it aebarycen}.
Then 0.2-12\,keV light curves with 1\,s binning (the minimum time resolution
of the XIS 1/8 window mode)
were extracted using the source and background
regions for the spectra.
To find the most reliable period, timing analyses were performed
by {\it powspec} and {\it efsearch}.
A folded light curve was made with the best period by {\it efold}.

The latest calibration database (CALDB:\,20080602)
was applied to the unfiltered HXD event data using {\it hxdpi} and {\it hxdgrade}.
The PIN and GSO event data were extracted from the
newly calibrated data with basic
criteria\footnote{The HXD basic criteria were taken from http://www.astro.isas.ac.jp/suzaku/analysis/7step\_HXD\_20080501.txt.}
using {\it xselect}.
The tuned background event data (bgd\_d) were utilized to estimate
the non X-ray background (NXB).
The cosmic X-ray background (CXB) was estimated
based on the HEAO-1 result \citep{boldt1987}.
Then the good time interval (GTI) files were generated
by merging the PIN/GSO GTI extension and the bgd\_d GTI extension using {\it mgtime}.
Light curves and spectra were extracted using {\it xselect}.
For the source spectra, the dead time corrections were applied using {\it hxddtcor}.
The modeled background event data (bgd\_d) were utilized in our analyses.
The spectra were binned to at least 30 counts in each spectral bin by {\it grppha}.

Since both SGRs 1900$+$14 and 1806$-$20 are located in the galactic plane,
the galactic ridge emission (GRE) should be considered as well as NXB and CXB.
It is not easy to estimate the contribution of the GRE, so we used
the HESS\,J1804$-$216\_BGD data \citep{bamba2007} to
estimate both the CXB and GRE for the SGR\,1806$-$20 observation.
Note that all PIN diodes were operated
at 500\,V during the
HESS\,J1804$-$216\_BGD observation,
while half of the PIN diodes (UNITID$\lesssim$7) were operated
at 400\,V during the
SGR\,1806$-$20 observation.
Since simultaneous analysis of these two observations
with different modes is not available,
we concentrate on the 500\,V data (i.e., UNITID$>$7).

\subsection{Spectral Analysis}\label{spc_ana}
Since a two blackbody function (2BB) and a blackbody plus a power law model (BB$+$PL)
have been suggested for the quiescent SGR X-ray spectra,
we used them for our spectral analyses.
An absorption model was applied to both models.

The XIS spectra of SGR\,1900$+$14
are acceptably fitted by 2BB and BB$+$PL, for
which the spectral parameters are summarized in table \ref{tab:spc_summary},
and the spectra are shown in figure \ref{fig:spc_summary} (a).

The non-thermal hard X-ray emission component discovered by INTEGRAL \citep{molkov2005}
was detected by the PIN diodes for SGR\,1806$-$20.
The emission is significant even if we increase
the background flux by a factor of 1.02 (corresponding
to the current background fluctuation of the PIN diodes).
Since there are not enough statistics to investigate the spectral shape
of the non-thermal hard emission,
we performed a spectral analysis using a power law model (HPL)
with a fixed index of $\Gamma = 1.6$ as measured by INTEGRAL.
The 10-70\,keV flux was $F = 2.9_{-1.2}^{+0.9} \times 10^{-11}$\,erg\,cm$^{-2}$\,s$^{-1}$,
where the quoted errors are 68\% confidence levels.
The unabsorbed 1-200\,keV flux was estimated to be
$F \sim 4.3 \times 10^{-11}$\,erg\,cm$^{-2}$\,s$^{-1}$
using the best-fit spectral parameters
which is lower by a factor of $\sim3$ than the flux
observed by INTEGRAL ($F \sim 1.3 \times 10^{-10}$\,erg\,cm$^{-2}$\,s$^{-1}$; \cite{molkov2005}).
Then we performed the broadband spectral analyses of the XIS
and the PIN diodes for SGR\,1806$-$20.
The spectra are well described by 2BB plus HPL (2BB$+$HPL)
or BB$+$PL plus HPL (BB$+$PL$+$HPL).
The index and flux of the non-thermal hard emission are
fixed to $\Gamma = 1.6$ and $F = 2.9 \times 10^{-11}$\,erg\,cm$^{-2}$\,s$^{-1}$, respectively.
The broadband spectra of SGR\,1806$-$20 are shown
in figure \ref{fig:spc_summary} (b).

\subsection{Timing Analysis}
\begin{figure}
  \begin{center}
    \FigureFile(160mm,80mm){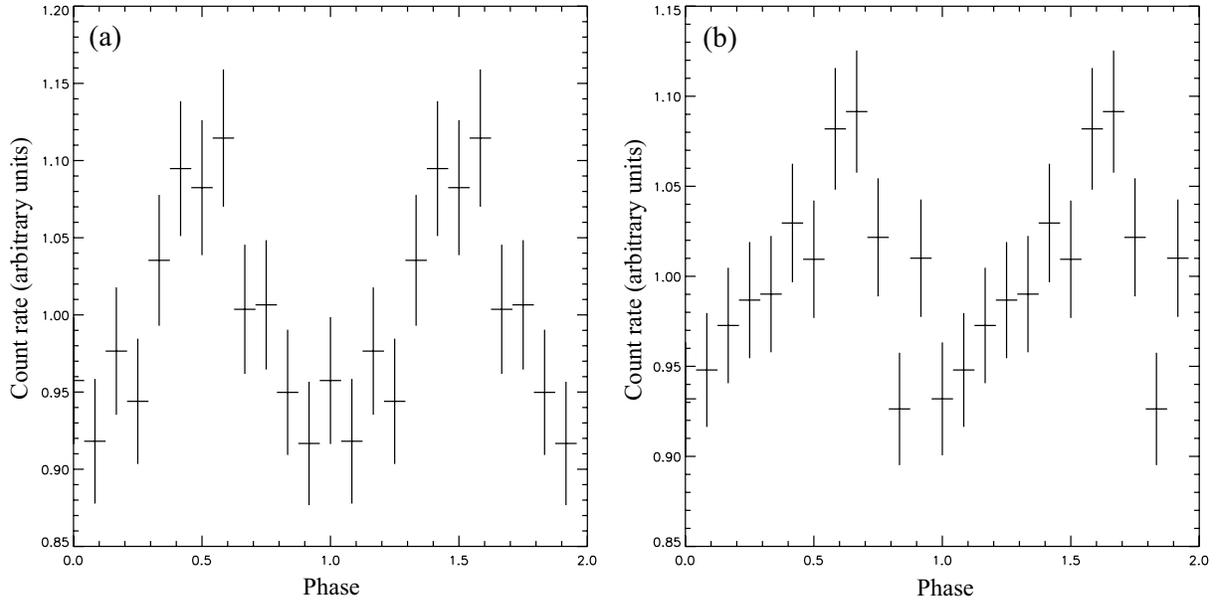}
  \end{center}
  \caption{Folded 0.2-12\,keV light curves for SGR\,1900$+$14 (a) and SGR\,1806$-$20 (b).}\label{fig:efold_summary}
\end{figure}

The pulse period of SGR\,1900$+$14 was searched for using
the XIS 0.2-12\,keV light curves.
Two candidates with nearly the same confidence level
were found from an epoch-folding analysis: $P = 5.1998\pm0.0002$\,s ($\chi^2 = 24.8$)
and $P = 5.2043\pm0.0003$\,s ($\chi^2 = 24.7$).
The folded light curve with the first period shows a pulsation
(the left panel in figure \ref{fig:efold_summary}),
while the other case does not display a clear pulsation (not shown in figure \ref{fig:efold_summary}).
In addition, the first pulse period is consistent with
that determined by XMM-{\it Newton} on April 1 2006 \citep{mereghetti2006},
the same date as that of the {\it Suzaku} observation.
The most plausible pulse period is therefore $P = 5.1998\pm0.0002$\,s.
The pulse period derivative between the XMM-{\it Newton} observation in September 2005
($P = 5.198346\pm0.000003$\,s; \cite{mereghetti2006})
and the {\it Suzaku} observation in April 2006 is estimated to be
$\dot{P} = (8.7\pm1.2) \times 10^{-11}$\,s\,s$^{-1}$.
This is lower by a factor of 0.8
than the first measurement $\sim100$ days prior to the giant flare
on 27 August 1998 \citep{kouveliotou1999}.
Using the folded light curve with the best period, the pulsed fraction is found to be
$P_{\rm f} = 16\pm3$\%.
Here $P_{\rm f} = (C_{\rm max} - C_{\rm min})/(C_{\rm max} + C_{\rm min})$,
where $C_{\rm max}$ and $C_{\rm min}$ are the measured count rates at
the maximum and at the minimum, respectively.

The pulse period of SGR\,1806$-$20 from an epoch-folding analysis
is found to be $P = 7.6022\pm0.0007$\,s ($\chi^2 = 53.1$)
using XIS 0.2-12\,keV light curves.
The folded light curve displays the pulsations, as shown in the left panel
in figure \ref{fig:efold_summary}.
Comparing this with the reported pulse period derived
from the XMM-{\it Newton} observation on September 10 2006
($P = 7.5891\pm0.0002$\,s; \cite{esposito2007}),
the pulse period derivative is estimated to be $\dot{P} = (7.5\pm0.4) \times 10^{-10}$\,s\,s$^{-1}$.
This is consistent with previously reported values
derived from observations after the giant flare on 27 December 2004
(e.g., \cite{woods2007}).
Using the folded light curve, the pulsed fraction is found to be $P_{\rm f} = 8\pm2$\%.

%%%%%%%%%%%%%%%%%%%%%%%%%%%%%%%%%%%%%%%
\section{Discussion and Conclusion}
\begin{figure}
  \begin{center}
    \FigureFile(160mm,80mm){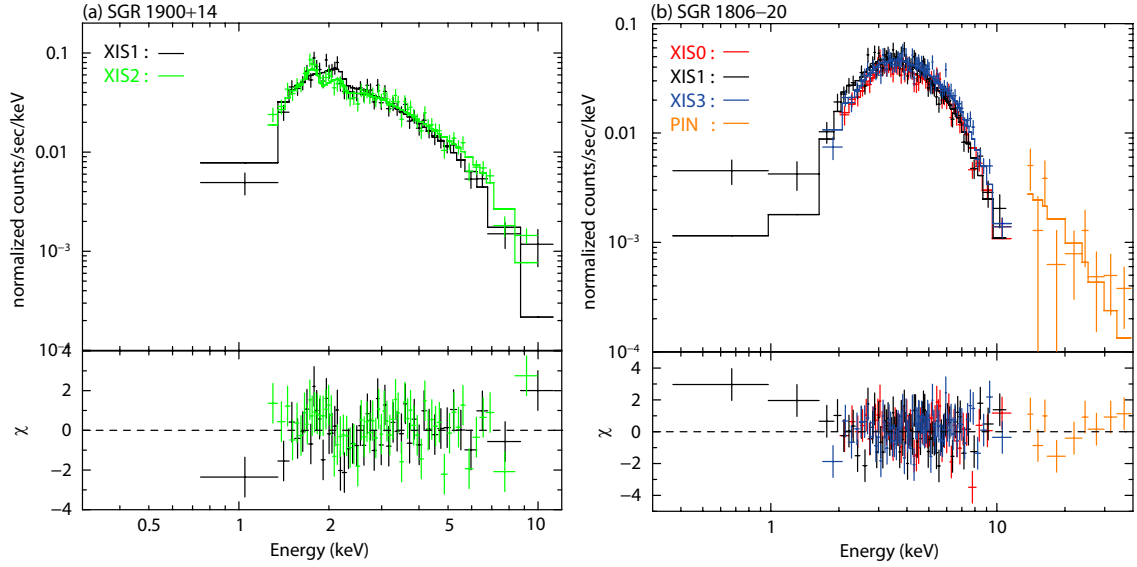}
  \end{center}
  \caption{Quiescent emission spectra of SGR\,1900$+$14 (a) and SGR\,1806$-$20 (b).}\label{fig:spc_summary}
\end{figure}

The quiescent emission spectra of the SGRs 1900$+$14 and 1806$-$20
measured by the XIS on-board {\it Suzaku} 
are well described by either a two blackbody function (2BB)
or a blackbody plus a power law model (BB$+$PL).
The 2BB temperatures ($kT_{\mathrm{LT}}$ and $kT_{\mathrm{HT}}$)
and radii ($R_{\mathrm{LT}}^2$ and $R_{\mathrm{HT}}^2$) show good agreement with
the $kT_{\mathrm{LT}}$-$kT_{\mathrm{HT}}$ and $R_{\mathrm{LT}}^2$-$R_{\mathrm{HT}}^2$
correlations reported by \citet{nakagawa2007b}.
A shallow dip is apparent at around 2.3 keV in the SGR\,1900$+$14
spectra as shown in figure \ref{fig:spc_summary} (a).
However, the XIS hardware team has cautioned that
the detector gain has some unresolved uncertainties around 2\,keV for
the window mode\footnote{The reports are available on http://www.astro.isas.jaxa.jp/suzaku/process/caveats/caveats\_xrtxis06.html.},
so this dip might not be real.
If we treat it as a line feature, however, it could be 
well represented by a Gaussian shape absorption with $E = 2.3\pm0.1$\,keV
and $\sigma = 0.2_{-0.1}^{+0.2}$\,keV, where $E$ is the line energy
and $\sigma$ is the line width.
If the dip is due to the fundamental absorption feature of proton cyclotron
resonant scatterings, the dipole magnetic field
is $B = 3.7\times10^{14}(E/2.3\,\mathrm{keV})$\,G.
We conducted an absorption line search for both SGRs 1900$+$14 and
1806$-$20 using a narrow Gaussian absorption line model
with the line center as a free parameter ranging from 0.2\,keV to 12\,keV.
No feature was found except for the above mentioned $E \sim 2.3$\,keV
shallow dip of SGR\,1900$+$14.
The upper limits of the feature are $\tau < 0.3$ (SGR\,1900$+$14)
and $\tau < 0.4$ (SGR\,1806$-$20).

The non-thermal hard X-ray emission of SGR\,1806$-$20, discovered by INTEGRAL
in a 1.6\,Ms long observation,
was detected by the PIN diodes on-board {\it Suzaku}
with a short exposure time ($\sim20$\,ks) because of
the good sensitivity and the very low background level of this instrument.

\begin{figure}
  \begin{center}
    \FigureFile(160mm,160mm){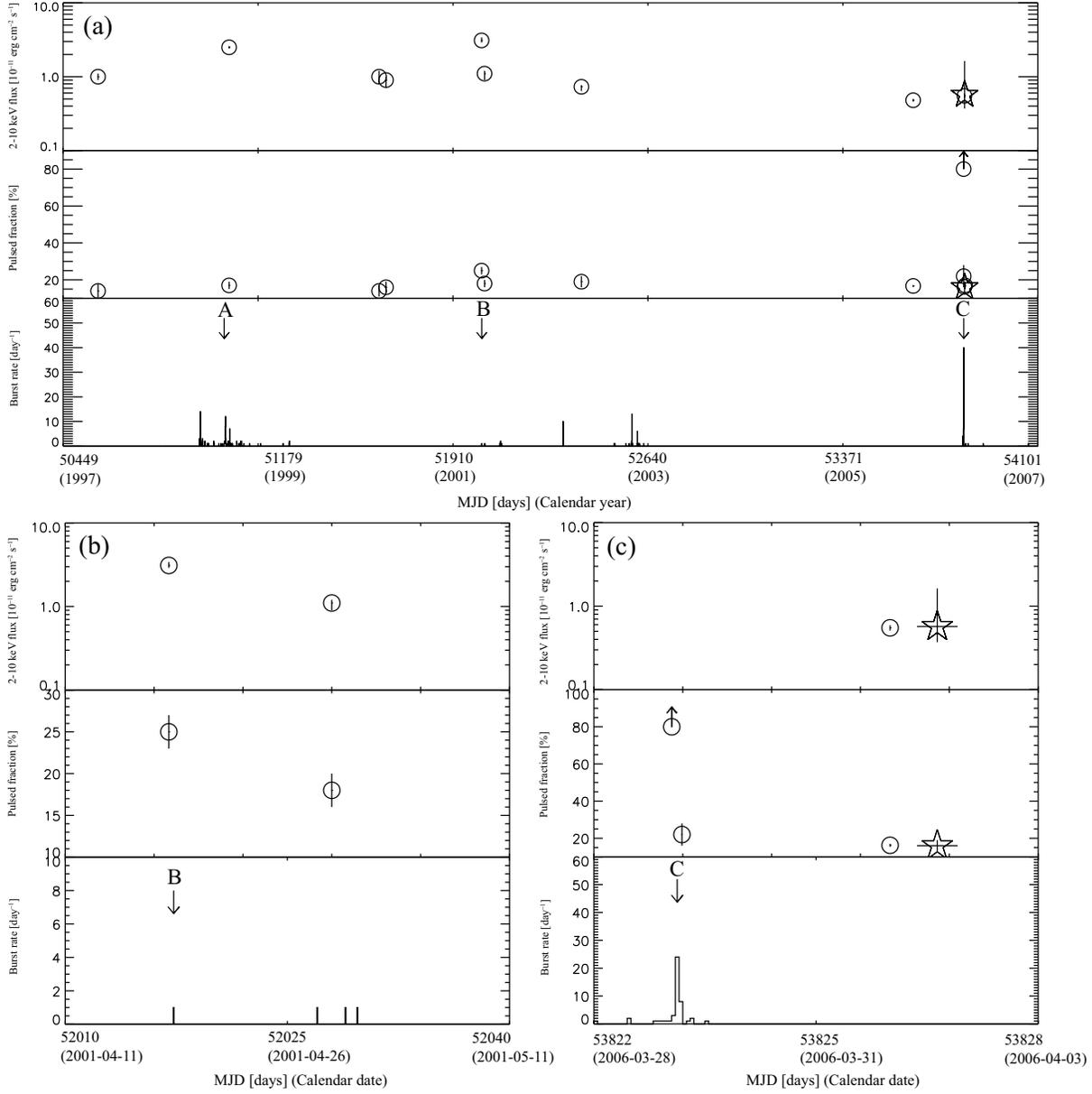}
  \end{center}
  \caption{Panel (a): time variations of the 2-10\,keV flux, the pulsed fraction and the burst rate
  from 1997 to 2007, where A, B and C indicate the days of the giant flare on 27 August 1998,
  the intermediate flare on 18 April 2001 and the extraordinary
  burst activity with more than 40 bursts on 29 March 2006, respectively.
  Panels (b) and (c): expanded plots around the days of B and C, respectively.
   The {\it Suzaku} observation is plotted with star symbols.}\label{fig:1900_tvar}
\end{figure}

The periodicities of SGRs 1900$+$14 and 1806$-$20 (table \ref{tab:timing_summary})
are detected despite the short exposure time ($\sim20$\,ks),
again because of the low background level of the {\it Suzaku} XIS.
It has been recognized that the pulsed fractions of SGRs
are not stable, and that their pulse shapes are complicated.
One example is the fact that the pulsed fraction displayed
a clear drop-off around the day of the giant flare from SGR\,1806$-$20
on 27 December 2004 \citep{woods2007}.
To perform a deeper study of the pulse characteristics of SGR\,1900$+$14,
we compared the pulsed fractions to the X-ray fluxes and burst rate.
The pulsed fractions and the X-ray fluxes have been derived both from
our work (tables \ref{tab:timing_summary} and \ref{tab:spc_summary})
and from the literature \citep{mereghetti2006, esposito2007, israel2008},
while the burst rate was estimated from the list
on the IPN web site\footnote{The burst list is available on
http://www.ssl.berkeley.edu/ipn3/sgrlist.txt.}.
Panel (a) in figure \ref{fig:1900_tvar} shows
their time histories over 10 years from 1997 to 2007,
where A, B and C indicate the days of
the giant flare on 27 August 1998, the intermediate flare on
18 April 2001 and the extraordinary burst activity with
more than 40 bursts on 29 March 2006, respectively.
The results of the {\it Suzaku} observation are plotted with
star symbols.

There seems to be drop-off of the pulsed fraction around
the days of B and C.
Panels (b) and (c) in figure \ref{fig:1900_tvar}
give an expanded view of panel (a) for days B and C, respectively.
In panels (b) and (c),
the pulsed fractions display a clear drop-off after
both active days (i.e., B and C).
No burst activity was found after the day B for 9 days,
and then the quiet burst activity appeared.
Since the day C, the burst activity had been gradually disappearing.
No bursts were reported by the IPN during the {\it Suzaku} observation
(the star symbol in figure \ref{fig:1900_tvar}).
The known bursts before and after the {\it Suzaku} observation
were detected at 12:19:51 on 29 March 2006
and at 00:23:48 on 5 April 2006, respectively.
On the contrary, SGR\,1806$-$20 displayed
one burst during the {\it Suzaku} observation,
and moderate, steady burst activity before and after it.
This drop-off is found in response not only to
the giant flare but also to the extraordinary burst activity.
(This implies that the pulsed fraction is
related to burst activity.)
One possible explanation is to assume that magnetic confinement
stores a vast amount of energy in the neutron star atmosphere or
magnetosphere
through energy injections such as starquakes,
and that the pulsed fraction is then increased.
After that the stored energy may be released as bursts
either through reconnection, or
some sort of instability such as the thermonuclear
explosions of X-ray bursts.
A recent theoretical study reports that
a variety of pulse morphologies can be
caused by trapped fireballs released by starquakes \citep{jia2008}.

We conclude by pointing out that monitoring the time variations
of the pulsed fraction and
the burst rate is an important element in revealing the burst mechanism.
This will be achieved by wide field detectors such as
the Monitor of All-sky X-ray Image (MAXI) on-board
the international space station \citep{matsuoka1997}.

%%%%%%%%%%%%%%%%%%%%%%%%%%%%%%%%%%%%%%%
% Acknowledgements
%%%%%%%%%%%%%%%%%%%%%%%%%%%%%%%%%%%%%%%
\bigskip
We would like to thank K. Hurley for helpful comments
and suggestions to improve our paper.
We are also grateful to the XIS team for useful discussions
on the timing analyses and the calibrations of the window mode.
This work is supported in part by a special postdoctoral researchers
program in RIKEN.
Y.E.N. is supported by the JSPS Research Fellowships for Young Scientists.

\onecolumn
%%%%%%%%%%%%%%%%%%%%%%%%%%%%%%%%%%%%%%%
% Figures
%%%%%%%%%%%%%%%%%%%%%%%%%%%%%%%%%%%%%%%

%%%%%%%%%%%%%%%%%%%%%%%%%%%%%%%%%%%%%%%
% Tables
%%%%%%%%%%%%%%%%%%%%%%%%%%%%%%%%%%%%%%%
\begin{table}
  \caption{A summary of {\it Suzaku} ToO observations.}\label{tab:obs_summary}
  \begin{center}
    \begin{tabular}{llllllll}
     \hline\hline
     SGR          & SeqNum\footnotemark[$*$] & \multicolumn{2}{l}{Observation Date (MJD)} & $T$\footnotemark[$\dagger$] & XIS\footnotemark[$\ddagger$] & \multicolumn{2}{l}{Regions\footnotemark[$\S$]} \\
              &                          & Start     & End                            & (ks)         &   & FI  & BI \\
     \hline
     1900$+$14  & 401022010 & 53826.363 & 53826.911 & 17.0 & 1, 2 & $\timeform{2.3'}\times\timeform{5.0'}/\timeform{2.3'}\times\timeform{6.7'}$ & $\timeform{2.3'}\times\timeform{7.0'}/\timeform{2.3'}\times\timeform{3.2'}$ \\
     1806$-$20  & 401021010 & 54189.631 & 54190.063 & 19.6 & 0, 1, 3 & $\timeform{2.2'}\times\timeform{5.3'}/\timeform{2.2'}\times\timeform{3.7'}$ & $\timeform{2.2'}\times\timeform{7.0'}/\timeform{2.2'}\times\timeform{3.5'}$ \\
     \hline
     \multicolumn{8}{@{}l@{}}{\hbox to 0pt{\parbox{180mm}{\footnotesize
       \footnotemark[$*$] {\it Suzaku} sequence number.
       \par\noindent
       \footnotemark[$\dagger$] Net exposure time for each observation. 
       \par\noindent
       \footnotemark[$\ddagger$] XIS sensor number.
       \par\noindent
       \footnotemark[$\S$] Extracted source/background regions for front-illuminated (FI) and back-illuminated (BI) CCDs.
     }\hss}}
    \end{tabular}
  \end{center}
\end{table}

\begin{table}
 \small
  \caption{A summary of timing analyses.}\label{tab:timing_summary}
  \begin{center}
    \begin{tabular}{llll}
     \hline\hline
     SGR          & Epoch (MJD TDB)    & $P$\footnotemark[$*$] & $\dot{P}$\footnotemark[$\dagger$]   \\
     \hline
     1900$+$14  & 53826.001  & $5.1998\pm0.0002$ & $8.7\pm1.2$ \\
     1806$-$20  & 54189.0009 & $7.6022\pm0.0007$ & $75\pm4$ \\
     \hline
     \multicolumn{4}{@{}l@{}}{\hbox to 0pt{\parbox{85mm}{\footnotesize
       \footnotemark[$*$] Pulse periods in units of s.
       \par\noindent
       \footnotemark[$\dagger$] Pulse period derivatives in units of $10^{-11}$\,s\,s$^{-1}$.
     }\hss}}
    \end{tabular}
  \end{center}
\end{table}

\begin{table}
 \small
  \caption{A summary of spectral parameters.
  The XIS spectra of SGR\,1900$+$14 are fitted with 2BB or BB$+$PL,
  while the XIS and PIN diodes spectra of SGR\,1806$-$20 are fitted
  with 2BB$+$HPL or BB$+$PL$+$HPL. The spectral parameters of
  the non-thermal hard emission of SGR\,1806$-$20 are not shown
  in the table because the index and the flux were fixed to $\Gamma = 1.6$
  and $F = 2.9 \times 10^{-11}$\,erg\,cm$^{-2}$\,s$^{-1}$ (see subsection \ref{spc_ana}), respectively.}\label{tab:spc_summary}
  \begin{center}
    \begin{tabular}{llllllllll}
     \hline\hline
     SGR & Model & $N_{\mathrm{H}}$\footnotemark[$*$] & $kT_{\mathrm{LT}}$ ($kT_{\mathrm{BB}}$)\footnotemark[$\dagger$] & $R_{\mathrm{LT}}$ ($R_{\mathrm{BB}}$)\footnotemark[$\ddagger$] & $kT_{\mathrm{HT}}$\footnotemark[$\dagger$] & $R_{\mathrm{HT}}$\footnotemark[$\ddagger$] & $\Gamma$\footnotemark[$\S$] & $F$\footnotemark[$\|$] & $\chi^2$ (d.o.f.) \\
      &      & ($10^{22}$\,cm$^{-2}$) & (keV) & (km) & (keV) & (km) & & & \\
     \hline
     1900$+$14 & 2BB      & $2.2\pm0.3$ & $0.49\pm0.06$ & $4.6_{-1.0}^{+1.7}$ & $1.5_{-0.2}^{+0.3}$ & $0.36_{-0.11}^{+0.16}$ & $\cdot\cdot\cdot$ & $4.1\pm0.5$ & 115 (105) \\
       & BB$+$PL  & $3.2_{-0.7}^{+0.9}$ & $0.29_{-0.10}^{+0.17}$ & $12.3_{-9.3}^{+71.8}$ & $\cdot\cdot\cdot$ & $\cdot\cdot\cdot$ & $2.8_{-0.5}^{+0.3}$ & $4.1\pm0.5$ & 106 (105) \\
     1806$-$20 & 2BB$+$HPL      & $7.8_{-0.9}^{+1.2}$ & $0.54_{-0.09}^{+0.08}$ & $3.1_{-1.3}^{+2.7}$ & $2.7_{-1.1}^{+3.4}$ & $0.06_{-0.04}^{+0.19}$ & $\cdot\cdot\cdot$ & $9.9\pm1.6$ & 196 (192) \\
       & BB$+$PL$+$HPL  & $8.0_{-1.1}^{+2.0}$ & $0.51_{-0.13}^{+0.09}$ & $3.6_{-1.6}^{+5.9}$ & $\cdot\cdot\cdot$ & $\cdot\cdot\cdot$ & $1.6_{-1.1}^{+2.2}$ & $9.9\pm1.6$ & 196 (192) \\
     \hline
     \multicolumn{10}{@{}l@{}}{\hbox to 0pt{\parbox{180mm}{\footnotesize
       \footnotemark[$*$] $N_{\mathrm{H}}$ denotes the photoelectric absorption with 90\% confidence level errors.
       \par\noindent
       \footnotemark[$\dagger$] $kT_{\mathrm{LT}}$, $kT_{\mathrm{HT}}$ and $kT_{\mathrm{BB}}$ denote blackbody temperatures with 90\% confidence level errors.
       \par\noindent
       \footnotemark[$\ddagger$] $R_{\mathrm{LT}}$, $R_{\mathrm{HT}}$ and $R_{\mathrm{BB}}$ denote the emission radius with 90\% confidence level errors.
       \par\noindent
       \footnotemark[$\S$] $\Gamma$ denotes the power law index with 90\% confidence level errors.
       \par\noindent
       \footnotemark[$\|$] $F$ denotes the 2-10\,keV flux in units of $10^{-12}$\,erg\,cm$^{-2}$\,s$^{-1}$ with 68\% confidence level errors.
     }\hss}}
    \end{tabular}
  \end{center}
\end{table}

\clearpage
%%%
% See the manual for the detail.
%%%

\end{document}